\documentclass[aps,prx,twocolumn,colorlinks=true,linkcolor=black,urlcolor=blue,citecolor=black,pdfhttps://www.overleaf.com/project/624ef482c726789d7c1a8f5cencoding=auto,longbibliography,superscriptaddress]{revtex4-2}
\usepackage{graphicx}
\usepackage{color}
\usepackage{amsmath}
\usepackage{amsfonts}
\usepackage{bbold}
\usepackage{amssymb}
\usepackage{tikz}
\usepackage{bm}
\usepackage{braket}
\usepackage{pgfplots}
\usepackage{dsfont}
\usepackage{comment}
\usepackage[colorlinks=true,
            citecolor=blue,
            pdfencoding=auto]{hyperref}
\usepackage{blindtext}
\usepackage{mathrsfs}
\usepackage[scr=boondoxo,scrscaled=1.05]{mathalfa}
\usepackage{lineno}
\usepackage{dirtytalk}

\makeatletter
\newcommand{\overleftrightsmallarrow}{\mathpalette{\overarrowsmall@\leftrightarrowfill@}}
\newcommand{\overrightsmallarrow}{\mathpalette{\overarrowsmall@\rightarrowfill@}}
\newcommand{\overleftsmallarrow}{\mathpalette{\overarrowsmall@\leftarrowfill@}}
\newcommand{\overarrowsmall@}[3]{%
  \vbox{%
    \ialign{%
      ##\crcr
      #1{\smaller@style{#2}}\crcr
      \noalign{\nointerlineskip}%
      $\m@th\hfil#2#3\hfil$\crcr
    }%
  }%
}
\def\smaller@style#1{%
  \ifx#1\displaystyle\scriptstyle\else
    \ifx#1\textstyle\scriptstyle\else
      \scriptscriptstyle
    \fi
  \fi
}
\makeatother

\begin{document}

\title{Local Hall Conductivity in Disordered Topological Insulators}


\author{Zachariah Addison}%
\affiliation{Department of Physics and Astronomy,
Wellesley College,
Wellesley, MA 02482, USA}
\author{Nandini Trivedi}
\affiliation{Department of Physics,
Ohio State University,
Columbus, OH 43210, USA}

\date{\today}

\begin{abstract}
We derive the expression for the local Hall conductivity for systems that lack translation symmetry and use it to study the local fluctuations of the Hall signal around disordered patches in magnetic insulators. 
We find that the regime in parameter space over which the system is a Chern insulating state {\em increases} upon inclusion of non-magnetic potential disorder. In addition, the phase space over which the topological Anderson insulator exists can be enhanced by breaking up a single disordered patch into multiple smaller patches with the same total amount of disorder. We expect our results will motivate the next generation of local scanning and local impedance spectroscopy experiments to visualize Hall currents around patches in the bulk of a disordered topological insulator.

\end{abstract}

\maketitle\

\section{Introduction}

Topological insulators are usually theoretically studied in the context of non-interacting infinite periodic systems.  For these systems the eigenstates of the Hamiltonian are single particle Bloch states indexed by a crystal momentum taking values in the first Brillouin zone.  Topological phases of matter are then distinguished by topological invariants constructed as integral forms of the phase evolution of the Bloch states across the momentum space manifold.

 Real systems are finite in which the discrete translation symmetry is broken, either by the system's edge or by defects in the crystalline structure of the material.  However, topological phases continue to exist in experiments as seen in quantized conductances or topologically protected edge states.  Without translation symmetry there are only a few analytic and numerical techniques available to diagnose the usual topology of the system: among them is the Bott index \cite{resta1998quantum, prodan2010entanglement, hastings2011topological,huang2018quantum, lin2021real, benalcazar2022chiral}, local topological markers \cite{kitaev2006anyons,bianco2011mapping,kubota2017controlled,marrazzo2017locality,bourne2018non,li2019local,caio2019topological,varjas2020computation,sykes2021local,velury2021topological,mondragon2024robust}, and spectral localizers \cite{prodan2011disordered,loring2011disordered,loring2015k,loring2017finite,cerjan2022operator,liu2023mixed,ochkan2024non,chadha2024real,cerjan2024classifying}.

Here we focus on the Chern insulating phase that can be observed in experiments via measurements of the global Hall conductance or by the imaging of uni-directional edge states.  To connect with these experimental properties we focus not on the Chern number as defined by the integral of the Berry curvature over the first Brillouin zone \cite{thouless1982quantized} or related topological labels, but instead on the Hall conductivity, a quantity that can be calculated in the absence of crystallinity and that can also be directly measured in transport. 

In the context of translation symmetry breaking, there have been efforts recently to understand the local topological properties of systems via topological markers and spectral localizers that have well-defined local counterparts, unlike the Bott index \cite{cerjan2024classifying}. 
A local Chern number can be calculated by integrating the local Chern marker over a finite region in the bulk of the sample and neglecting contributions from the edge of the system.  This requires defining a suitable area of integration whose asymptotic convergence to the bulk integer-valued Chern number can be model or parameter dependent.  The integral of the local Chern marker over the entire finite system is equal to zero. To circumvent this issue, periodic boundary conditions can be applied on the system, however, the canonical form of the local Chern marker written in terms of position space operators requires some care in defining the local marker for periodic systems \cite{bianco2011mapping,wheeler2019many,bednik2024multifractaility}.

Motivated by recent progress in the experimental measurement of the local optical and transport properties of topological materials \cite{pascher2014imaging,lachman2015visualization,jiang2017plasmon,  kim2019observation,ferguson2023direct,ji2024local}, we recast questions about the local topological properties of materials in terms of quantities that are experimentally measurable, and specifically focus on the local Hall conductivity.  

The local Hall conductivity describes local charge currents that flow transverse to an applied voltage  and can be evaluated in the bulk of the sample without reference to the global geometry of the system (i.e. whether or not the system has an edge or has periodic boundary conditions).   Much attention has been paid to local currents at the boundary between different topological phases, usually between trivial and topologically non-trivial phases, where it has been shown that these boundaries possess robust topological edge modes that are insensitive to most small perturbations of the system and can carry a quantized conductance \cite{halperin1982quantized, buttiker1988absence, teo2010topological}.  While several local scanning probes are available, such as scanning potential microscopy \cite{mccormick1999scanned}, scanning single electron transistor \cite{yacoby1999electrical}, subsurface charge accumulation imaging \cite{finkelstein2000topographic}, and scanning gate microscopy \cite{paradiso2012imaging}, they have not been useful to study the bulk of the pristine topological insulator since most require a finite local density of states which vanishes in the bulk of an insulator.

  However, in principle the local Hall conductivity is still an observable and progress is being made toward its direct measurement. 
  We therefore highlight the bulk contributions to currents by investigating the fluctuations of the local Hall conductivity in a system with local translation symmetry breaking for different types of disorder. Previously the local Hall conductivity has been used as a cross-hair marker to reproduce the local Chern marker by appropriate summation over the bulk contribution in the neighborhood of the local conductivity~\cite{d2022quantized}.  

Here we provide an alternative derivation of the local Hall conductivity that can be applied in the context of periodic boundary conditions such that its trace reproduces the global Chern number.  Instead of using the continuity equation to determine the local current operator, we calculate the local conductivity by coupling the system to a local electromagnetic vector potential.  We then solve the von Neumann equation in the presence of spatially homogeneous electric fields to extract the local optical conductivity tensor.  From this we derive an analytic expression for the local Hall conductivity in a form congruent with the canonical expressions for linear and non-linear optical processes.  The form of our expressions parse paramagnetic and diamagnetic contributions to the local conductivity and makes direct connection to the canonical structure for the momentum space Berry curvature and global Hall conductivity.  We express the local conductivity via the expectation value of velocity operators that are well defined for both periodic and non-periodic boundary conditions. 

There are three contexts in which disorder can induce topological phases: (1) magnetic disorder induced band inversions \cite{qin2016disorder,mo2022imaginary,ling2024disorder,peng2024structural}, (2) structural amorphous disorder \cite{xiao2017photonic,mansha2017robust,chern2019topological, agarwala2019topological,marsal2020topological,corbae2023amorphous}, and (3) Anderson potential disorder \cite{li2009topological, groth2009theory,jiang2009numerical,prodan2010entanglement,xing2011topological,song2012dependence}.  In case (3), a topological phase can be induced by a disorder potential that couples to the local electronic density of the system.  These phases are usually studied in the context of a boxed disorder distribution where sites have an equal probability to be perturbed by a potential bounded by some energy scale $V_0$ \cite{li2009topological,groth2009theory,jiang2009numerical,guo2010topological,prodan2011disordered,prodan2010entanglement,prodan2011three,xing2011topological,ryu2012disorder,song2012dependence,cheng2023topological}.

Here we propose another mechanism in which disorder can drive a topological phase transition.  Starting with a magnetic insulator we add disorder that introduces semi-metallic patches into the system (akin to the addition of trivial substitutional atoms into the material). The disorder we study is different from previous work that has investigated the topological Anderson insulator phase space associated with the parameters of the boxed disorder distribution $V_0$ and the Fermi energy \cite{li2009topological,groth2009theory,jiang2009numerical,prodan2010entanglement,guo2010topological,xing2011topological}.

\medskip

\noindent Our main results are the following: 

\noindent (i) We find that semimetallic disorder patches can induce a topological phase transition from a trivial magnetic insulator to a topological Chern insulator. 

\noindent (ii) We show that the parameter space over which the topological phase exists, 
remarkably {\it increases} with increasing amount of disorder in the system.  

\noindent (iiii) We investigate a variety of semi-metallic patch configurations that can induce a topological Anderson transition. The correlation between the local Hall conductivity with the disorder potential allows us to determine the mechanism that drives the system to a globally topological or trivial state. 
From these insights we show 
that for a given total amount of disorder, delocalizing or spreading out the disorder puddles is advantageous to induce a transition into a topologically non-trivial phase. 

\section{Defining the Local Hall Conductivity}

We start by calculating the global Hall conductivity by coupling the system to a homogeneous electric field and finding the expectation value of the global current operator in the presence of the field.  We then generalize to a {\em local} conductivity by coupling the system to a local time-dependent electromagnetic vector potential to determine the local current operator. We then extract the zero frequency local Hall conductivity that determines the local susceptibility of the system to generate currents transverse to an applied homogeneous electric field in the presence of our chosen local disorder.


The 2D global Hall conductivity describes the component of the conductivity tensor $\overleftrightsmallarrow{\sigma}$ that is invariant under rotations about an axes perpendicular to the xy-plane: $\sigma_{H}=(\sigma_{xy}-\sigma_{yx})/2$.  It can be determined by calculating the current 

\begin{equation}
\bm{j}(t)=\text{Tr}[\widehat{\rho}(t)\widehat{\bm{j}}(t)]
\label{curr}
\end{equation}

\noindent
to first order with respect to a homogeneous electric field $\bm{E}$.  Here $\widehat{\bm{j}}(t)=-\bm{\nabla}_{\bm{A}(t)}\widehat{H}(t)/V$ is the current operator, $\widehat{H}(t)$ is the Hamiltonian in the presence of the electromagnetic vector potential $\bm{A}(t)$, $V$ is the volume of the system, and the state of the system is described by a density matrix $\widehat{\rho}(t)$ that satisfies the von Neumann equation $i\hbar \partial_t \widehat{\rho}(t)=[\widehat{H}(t),\widehat{\rho}(t)]$.  Here we neglect electron-electron interactions and thus can work with the single particle density of matrix that we express in terms of the basis of single particle eigenstates of the unperturbed Hamiltonian $\ket{n}$.  We expand equation \eqref{curr} to first order in the electric field to calculate the Hall conductivity.

The current and density operators can be expanded in a power series with respect to $\bm{A}(t)$.  We denote the Fourier transform of the components of an operator in the basis of eigenstates of $\widehat{H}(t)$ in the absence of $\bm{A}(t)$ as $\mathcal{O}_{nm}^{(l)}(\omega)=\bra{n}\widehat{\mathcal{O}}^{(l)}(\omega)\ket{m}$ where the superscript denotes the $l$-th order term in the power series expansion.  The current, $\bm{j}(\omega)$, to first order in $\bm{A}(\omega)$ takes the form

\begin{equation}
\bm{j}^{(1)}(\omega)=\sum_{n,m}\rho^{(1)}_{nm}(\omega)\bm{j}_{mn}^{(0)}(0)+\rho^{(0)}_{nm}(0)\bm{j}_{mn}^{(1)}(\omega)
\label{curr1}
\end{equation}

\noindent
where

\begin{align}
\bm{j}_{nm}^{(0)}(\omega)&=-\dfrac{1}{V}\bra{n} \bm{\nabla}_{\bm{A}(t)}\widehat{H}(t)\ket{m}\bigg|_{\bm{A}(t)=0} \delta_{\omega,0} \\
\bm{j}_{nm}^{(1)}(\omega)&=-\dfrac{1}{V}\sum_{i}\bra{n} \bm{\nabla}_{\bm{A}(t)}\partial_{A_i(t)}\widehat{H}(t)\ket{m}\bigg|_{\bm{A}(t)=0}A_i(\omega)
\end{align}

\noindent
and

\begin{align}
\rho_{nm}^{(0)}(\omega)&=\delta_{\omega,0}\delta_{nm}P_n \nonumber\\
\rho_{nm}^{(1)}(\omega)&=\dfrac{P_n-P_m}{E_n-E_m+\hbar\omega}e\bm{v}_{nm}\cdot \bm{A}(\omega)
\label{denmat}
\end{align}

\noindent
where $|P_l|^2$ denotes the occupation probability of eigenstate $\ket{l}$ in the unperturbed system, $\bm{v}_{nm}=-V\bm{j}_{nm}^{(0)}(0)/e$, and $e>0$ is the electric charge.  For simplicity in the following we will suppress the time dependence on quantities and drop the symbol $|_{\bm{A}(t)=0}$ assuming that in the following expressions that this limit is to be taken after any differentiation with respect to the electromagnetic vector potential.

Using the identities:

\begin{align}
e\bm{v}_{nm}&=\delta_{nm}\bm{\nabla}_{A}E_n(\bm{A})+(E_n-E_m)\bm{\nabla}_{A}\braket{n_{\bm{A}}|m} \nonumber\\
(M^{-1})_n^{pq}&=\partial_{A_p}\partial_{A_q}E_n[\bm{A}]/e^2-\sum_{m\neq n} \bigg[\dfrac{v_{nm}^pv_{mn}^q+v^q_{nm}v^p_{mn}}{E_n-E_m}\bigg]
\label{CId}
\end{align}

\noindent
where $(M^{-1})_n^{pq}=\bra{n}\partial_{A_p}\partial_{A_q}\widehat{H}\ket{n}/e^2$ is the inverse mass tensor and $\ket{n_{\bm{A}}}$ is the instantaneous eigenstate of the Hamiltonian in the presence of $\bm{A}(t)$.  The optical current to first order in the electric field can be written as

\begin{widetext}
\begin{equation}
\bm{j}^{(1)}(\omega)\cdot\bm{\hat{r}}_p=-\dfrac{e^2}{V}\sum_{j}\bigg[\sum_n \dfrac{P_n}{i\omega}\bigg(\partial_{A_j}\partial_{A_p}E_n[\bm{A}]/e^2\bigg)+\sum_{n\neq m}i\hbar\dfrac{(P_n-P_m) v^j_{nm}v_{mn}^p}{(E_n-E_m)(E_n-E_m+\hbar\omega)} \bigg]E^j(\omega)
\label{cond}
\end{equation}
\end{widetext}

\noindent
where $\hat{\bm{r}}_p$ is the unit vector in the $\bm{\hat{p}}$-direction.  The Hall conductivity $\sigma_H=(\sigma_{xy}(0)-\sigma_{yx}(0))/2$ is determined from the relation $\bm{j}^{(1)}(\omega)=\overleftrightsmallarrow{\sigma}(\omega)\bm{E}(\omega)$ and can be written as

\begin{equation}
\sigma_H=\dfrac{e^2}{V}\sum_{n\neq m}\dfrac{i\hbar}{2}\dfrac{(P_n-P_m)}{(E_n-E_m)^2} (\bm{v}_{nm}\times\bm{v}_{mn})\cdot\bm{\hat{z}}
\label{HallCon}
\end{equation}

\noindent
where $\widehat{v}^p$ denotes the component of the velocity operator in the $\hat{\bm{p}}$-direction.  For an insulating system at zero temperature $\sigma_H$ in the thermodynamic limit is integer valued in the absence of degenerate eigenvalues.  Here we will assume the disorder to be large enough that the energy eigenvalues $E_n$ can be taken to be quasi non-degenerate (i.e. all space-time symmetries of the system are microscopically broken).

Now turning to the local Hall conductivity, we first define a local conductivity, $\sigma_{ij}(\bm{r}_{pq})$, on a lattice by calculating the local current $\bm{j}(\bm{r}_{pq},t)=\text{Tr}[\widehat{\rho}(t)\widehat{\bm{j}}(\bm{r}_{pq},t)]$ to leading order in an applied homogeneous electric field.  The local current operator is defined as $\widehat{\bm{j}}(\bm{r}_{pq},t)=- \bm{\nabla}_{\bm{A}(\bm{r}_{pq},t)} \widehat{H}(t)/V$ where $\bm{A}(\bm{r}_{pq},t)$ is the value of the electromagnetic vector potential on the bond connecting lattice sites $\bm{r}_q$ and $\bm{r}_p$.  Using identities similar to the expressions in equation \eqref{CId} we find that the optical current to linear order in the electric field can be written as

\begin{widetext}
\begin{align}
\bm{j}^{(1)}(\bm{r}_{pq},\omega)\cdot\hat{\bm{r}}_i
=&-\dfrac{e^2}{V}\sum_{j}\bigg[\sum_n\dfrac{P_n}{ie\omega}\partial_{A_j}\bigg(\braket{n_{\bm{A}}|\bm{r}_p}\tilde{v}_{pq}^i\braket{\bm{r}_q|n_{\bm{A}}}\bigg)+\sum_{n\neq m}i\hbar\dfrac{(P_n-P_m) v^j_{nm}\Psi_m^*(\bm{r}_p)\tilde{v}_{pq}^i\Psi_n(\bm{r}_q)}{(E_n-E_m)(E_n-E_m+\hbar\omega)} \bigg] E^j(\omega)  
\end{align}
\end{widetext}

\noindent
where $\Psi_n(\bm{r}_p)=\braket{\bm{r}_p|n}$, and $\tilde{\bm{v}}_{pq}=\bra{\bm{r}_{p}}  \bm{\nabla}_{\bm{A}(t)}\widehat{H}(t)\ket{\bm{r}_q}/e$ is the local velocity operator (see the Appendix \ref{IdLoc} for details).  The real part of the local Hall conductivity $\sigma^{\text{local}}_H(\bm{r}_{pq})=(\sigma_{xy}(\bm{r}_{pq},0)-\sigma_{yx}(\bm{r}_{pq},0))/2$ takes the form

\begin{align}
    \sigma^{\text{loc}}_{H}(\bm{r}_{pq})&=\dfrac{e^2}{V}\sum_{n\neq m}\dfrac{i\hbar}{2}\dfrac{(P_n-P_m)\mathcal{P}_{nm}^{pq}}{(E_n-E_m)^2} (\tilde{\bm{v}}_{pq}\times\bm{v}_{mn})\cdot\bm{\hat{z}}
\end{align}

\noindent
where $\mathcal{P}_{nm}^{pq}=\Psi_n^*(\bm{r}_p)\Psi_m(\bm{r}_q)$ is the two-point overlap between eigenstates $n$ and $m$ of $\widehat{H}$.  In what follows we focus on the Hall bond conductivity we define as $\sigma_{H}(\bm{r}_{pq})=\sigma^{\text{local}}_{H}(\bm{r}_{pq})+\sigma^{\text{local}}_{H}(\bm{r}_{qp})$.

\section{Chern Insulating Model System}

To understand the role of disorder on the phase boundaries and local Hall conductivity in Chern insulating systems we adopt a minimal tight-binding model of nearest neighbor, $<i,j>$, hopping on the square lattice:

\begin{equation}
{H}= \sum_{<i,j>} {c}^\dagger_{i\sigma} T_{ij}^{\sigma\sigma'}{c}_{j\sigma'}
\end{equation}

\noindent
where $\widehat{c}^\dagger_{i\sigma}$ ($\widehat{c}_{i\sigma}$) creates (destroys) a fermion with spin $\sigma=\pm 1$ at lattice site $\bm{r}_i$ and $T_{ij}^{\sigma\sigma'}$ describes the coupling between sites.  In the absence of disorder we use a model for a massive Dirac fermion whose Bloch Hamiltonian takes the form

\begin{equation}
H(\bm{k})=\sum_i 2t\bigg(\sin(k_i a)\sigma_i-\cos(k_ia)\sigma_z\bigg)+\bm{M}\cdot\bm{\sigma}
\label{BlochHam}
\end{equation}

\noindent
where $t$ is the nearest neighbor hopping strength, $\bm{k}$ is the crystal momentum of the Bloch states $\ket{n,\bm{k}}$ taking values in the first Brillouin zone, and $n$ is a band index.  The mass of the fermion is parameterized by $\bm{M}$.

Since we want to explore the effects of on-site disorder and broken translation symmetry, we start by expressing $H(\bm{k})$ 
in the presence of disorder 
by using the position space basis as follows.  The coupling coefficients $T_{ij}^{\sigma\sigma'}$ are divided into two parts $T_{ij}^{\sigma\sigma'}=t_{ij}^{\sigma\sigma'}+\tau_{ij}^{\sigma\sigma'}$, where $t_{ij}^{\sigma\sigma'}$ has the property $t_{ij}^{\sigma\sigma'}=t^{\sigma\sigma'}(\bm{r}_i-\bm{r}_j)$, and $\tau_{ij}^{\sigma\sigma'}=\delta_{ij}\tau^{\sigma\sigma'}(\bm{r}_i)$ describes the effects from an inhomogeneous electrostatic disorder potential.  The couplings $t^{\sigma\sigma'}(\bm{r}_i-\bm{r}_j)$ of equation \eqref{BlochHam} can be expressed as

\begin{align}
t^{\sigma\sigma'}(\bm{r}_i-\bm{r}_j)&=t\bigg(\dfrac{i\bm{\sigma}_{\sigma\sigma'}\cdot(\bm{r}_i-\bm{r}_j)}{a}-\sigma^z_{\sigma\sigma'}\bigg)\nonumber\\
&+\delta_{ij}\bm{M}\cdot\bm{\sigma}_{\sigma\sigma'}\,,\,\text{for } |\bm{r}_i-\bm{r}_j|\leq a
\label{LocPerHop}
\end{align}

\begin{figure}[ht]
\includegraphics[width=.45\textwidth]{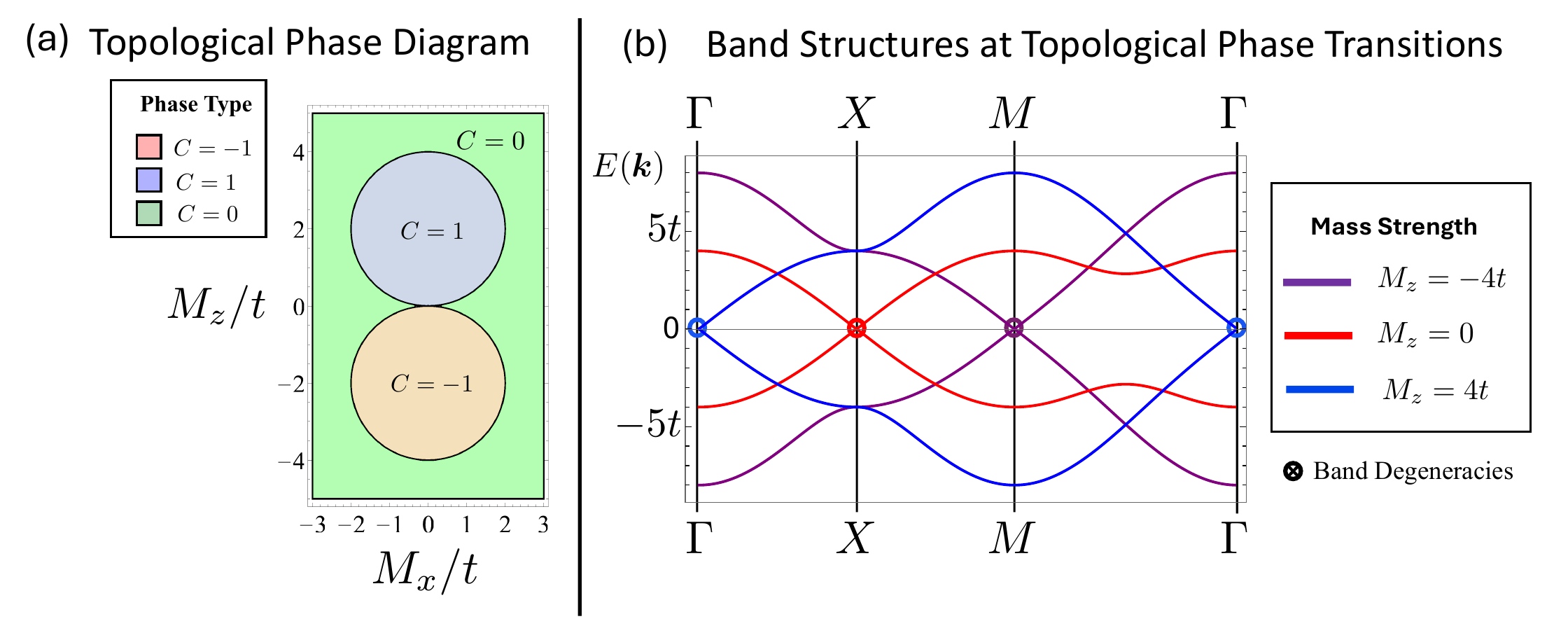}
\caption{{\bf Topological Transitions in a Clean System.} (a) At half filling the system is insulating and can be in three distinct topological states with $C=\pm1$ or $C=0$.  The phase boundaries as a function of $M_z/t$ and $M_x/t$ with $M_y=0$ are shown. (b) Band structure for critical $M_z$ on the phase-space line $M_x=M_y=0$.  At these values of $\bm{M}$ the lower and upper bands touch at high symmetry points in the Brillouin zone. 
}
 \label{fig1}
\end{figure}

In the absence of disorder, at half filling (i.e. $P_n=1$ for $1<n<N_s$) the system is insulating away from the surfaces

\begin{equation}
M_z/t=2\left(l_1+l_2\sqrt{1-\bigg(\dfrac{M_x}{2t}\bigg)^2-\bigg(\dfrac{M_y}{2t}\bigg)^2}\right)
\end{equation}

\noindent
with $l_1,l_2=\pm1$ and $M_x,M_y\leq 2t$,  where the energy gap closes at one or more of the high symmetry points in the Brillouin zone (see Fig. \ref{fig1}).

Figure \ref{fig1}(a) shows the topological phase diagram denoting regions of different Chern insulating phases of $H(\bm{k})$ as a function of $M_x/t$ and $M_z/t$ for $M_y=0$.  These distinct topological phases are distinguished by their zero temperature Hall conductivity.  In the presence of lattice translation symmetry equation \eqref{HallCon} takes its canonical form

\begin{equation}
\sigma_H=\dfrac{e^2}{\hbar}\dfrac{1}{V}\sum_{k,n}f_n(\bm{k})\Omega_n(\bm{k})
\end{equation}

\noindent
where $n$ is a band index, $f_n(\bm{k})=(1+e^{-\beta (E(\bm{k})-\mu)})^{-1}$ is the temperature, $k_BT=1/\beta$, and chemical potential, $\mu$, dependent Fermi distribution function, and $\Omega_n(\bm{k})=\hat{\bm{z}}\cdot \bm{\nabla}_k\times \bm{A}_n(\bm{k})$ and $\bm{A}(\bm{k})=\bra{n,\bm{k}}i\bm{\nabla}\ket{n,\bm{k}}$ are the Berry curvature and connection \cite{xiao2010berry}.
For insulating systems at $T=0$ the Hall conductivity is integer valued in units of $\sigma_0=e^2/h$: $\sigma_H=C\sigma_0$, with different topological phases distinguished by the value of the Chern number $C\in \mathbb{Z}$. Phase boundaries are marked in black and separate different Chern insulating phases.  Along these critical lines the energy gap closes at one of the high symmetry points in the Brillouin zone (see Fig. \ref{fig1}(b)).

\section{Hall conductivity, Local Currents, and Disorder}

In the presence of a disorder potential, lattice translation symmetry is broken and a crystal momentum $\bm{k}$ can no longer index eigenstates of $\widehat{H}$.  However, the Hall conductivity can still be calculated using equation \eqref{HallCon} with knowledge of the system's energy eigenstates and eigenvalues.

We first explore the role of disorder in a system with periodic boundary conditions with a single impurity patch centered at site $\bm{r}_0$ with potential $\tau^{\sigma\sigma'}(\bm{r}_i)=-\bm{M}\cdot\bm{\sigma}_{\sigma\sigma'}g_s(\bm{r}_i)$, where $g_s(\bm{r}_i)=e^{-|\bm{r}_i-\bm{r}_0|^2/2L^2}$.  The role of the potential is to locally eliminate the onsite mass term $\sim\delta_{ij}\bm{M}\cdot\bm{\sigma}_{\sigma\sigma'}$ in equation \eqref{LocPerHop}.  As $L$ increases the spatial extent of the disorder potential grows, locally eliminating the mass of a larger and larger number of lattice sites.

\begin{figure}
 \begin{centering}
\includegraphics[width=.45\textwidth]{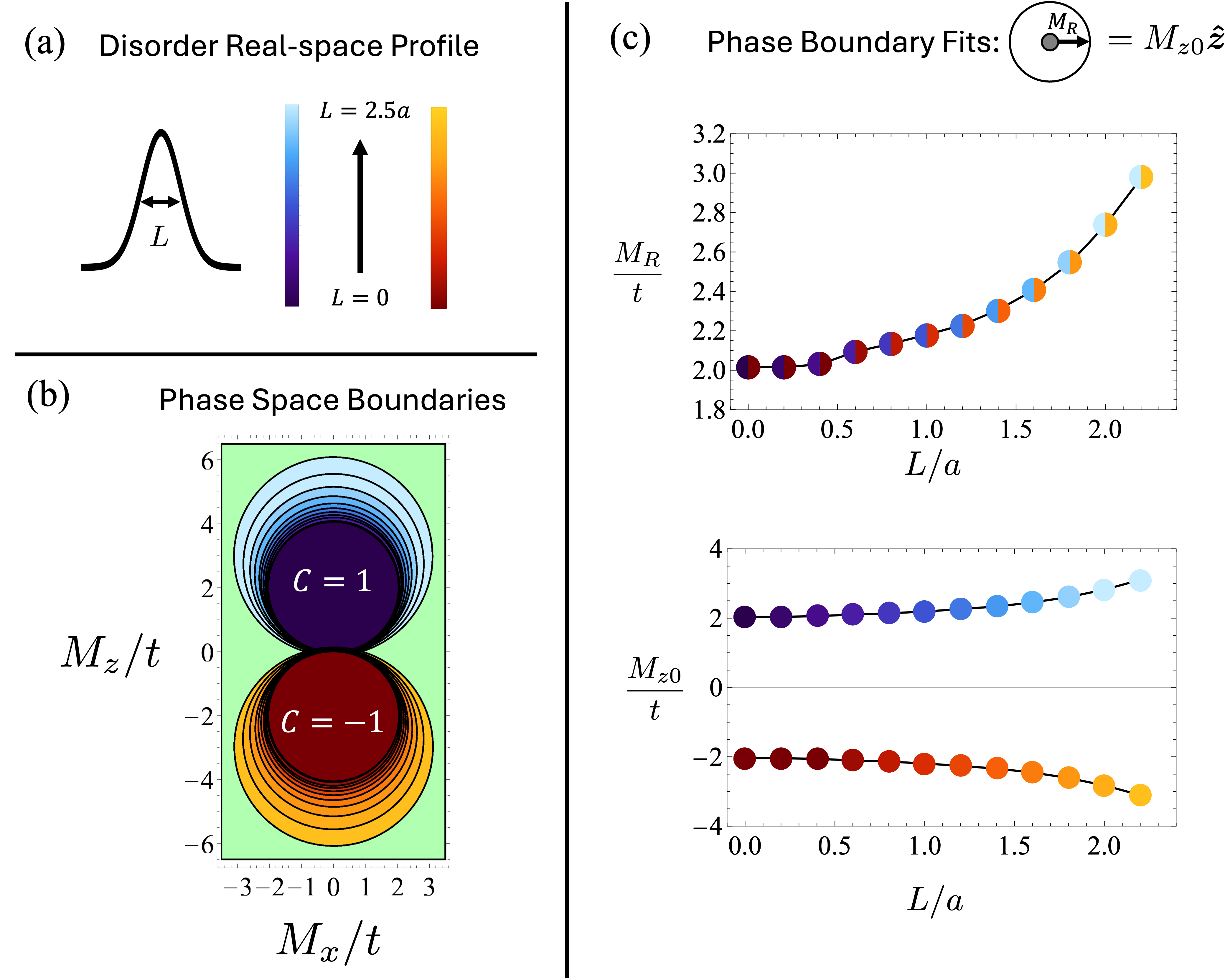}
\caption{{\bf Phase Diagram of Chern Insulator Model for Single Patch of Potential Disorder.} (a) Gaussian disorder profile on a typical patch length scale $L$ over which the system is semi-metallic and where the mass in equation \eqref{LocPerHop} vanishes. (b) The parameter region over which the topological phase exists {\em increases} with the size of disorder patch.  (c) Phase boundary remains circular, but the center $\bm{M}=M_{z0}\bm{\hat{z}}$ is shifted and radius $M_R$ increases non-linearly with the size of the disorder patch $L$.  Calculations were performed on a system of $N_s=400$ sites.
}
 \label{Boundaries1}
 \end{centering}
\end{figure}

Figure \ref{Boundaries1}(a) shows the gaussian profile of the disorder patch.  Figure \ref{Boundaries1}(b) plots the best fit topological phase boundaries in the $M_y=0$ plane as a function of $L/a$.  As $L$ is increased the phase boundaries remain circular with one edge of each circle pinned at the phase space point $\bm{M}=\bm{0}$. The radius of the circular non-trivial phase space volumes, $M_R$, increases non-linearly with $L$ (see Fig. \ref{Boundaries1}(c)).

\begin{figure*}
 \begin{centering}
\includegraphics[width=.85\textwidth]{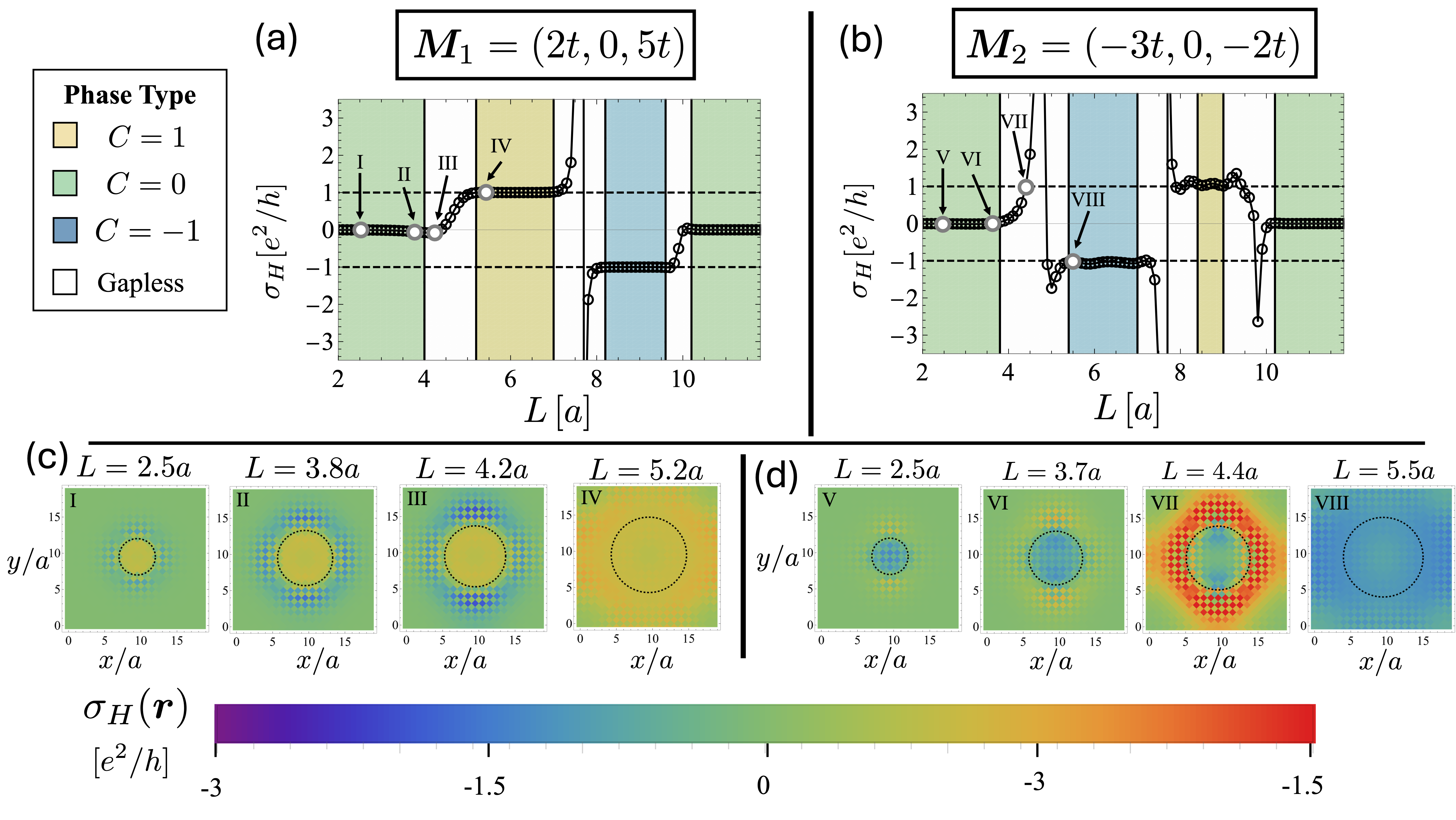}
\caption{{\bf Local Hall conductivity for Different Disorder Patch Sizes.} (a)-(b) 
 Hall conductivity calculated for two systems with $\bm{M}_1=2t\bm{\hat{x}}+5t\bm{\hat{z}}$ and $\bm{M}_2=-3t\bm{\hat{x}}+-2t\bm{\hat{z}}$ in the presence of a semimetallic disorder patch of size $L$. With increasing $L$ there are 
 topological phase transitions indicated by white regions that separate the colored regions with well defined Chern numbers;
here $N_s=400$.  
(c)-(d)  Local Hall conductivity maps for four different size patches in system $\bm{M}_1$ and system $\bm{M}_2$.  Contour where the gradient of the mass is peaked is shown by dashed black line. Topological phase transitions occur when the local Hall conductivity fluctuations permeate throughout the system.}
 \label{Boundaries2}
 \end{centering}
\end{figure*}

In the context of an experimental setup, the system's material parameters fix the value of the fermion mass.  In the clean limit this could lead to a topologically trivial ground state, however we see that doping the system with semi-metallic patches of disorder can induce a topological phase transition whereby a clean topologically trivial system can be induced into a topologically non-trivial Chern insulating state.

\begin{figure}
\includegraphics[width=.45\textwidth]{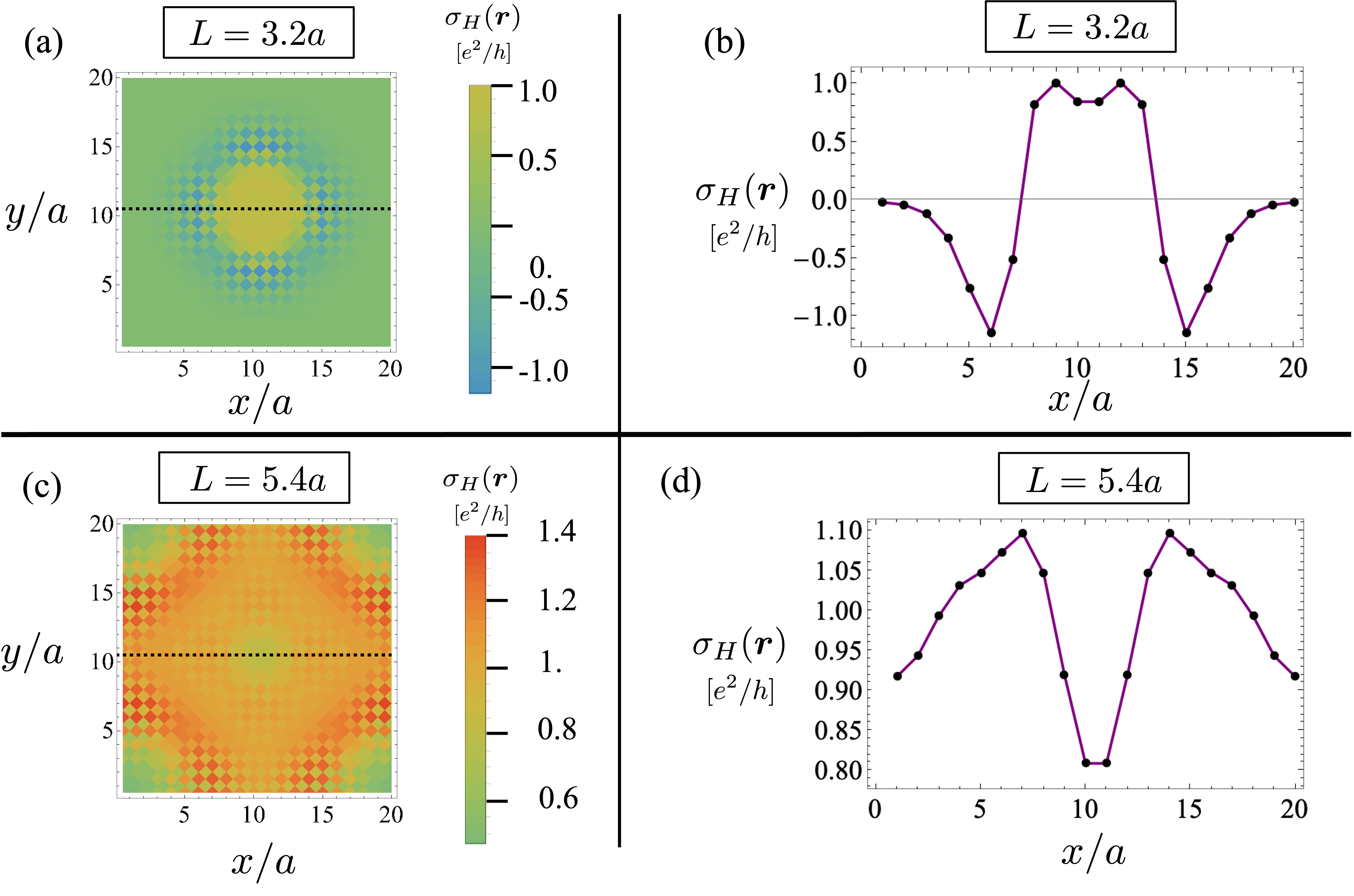}
\caption{{\bf Local Currents Across Disorder Patches.}  Local Hall conductivity for system with $\bm{M}=(2t,0,5t)$ for patch sizes $L=3.2a$ (a)-(b) and $L=5.4a$ (c)-(d).  (b) and (d) show a line cut ($y=10.5a$) across local Hall conductivity distribution showing how positive and negative fluctuations across sample average to zero in the trivial Chern insulating phase with $L=3.2a$ and average to $e^2/h$ in the $C=1$ phase with $L=5.4a$.}
 \label{LineCut}
\end{figure}

To explore the phase transition between trivial and non-trivial insulating states we consider two generic trivially insulating systems.  One system has a fermion mass of  $\bm{M}_1=2t\bm{\hat{x}}+5t\bm{\hat{z}}$ and the other $\bm{M}_2=-3t\bm{\hat{x}}+-2t\bm{\hat{z}}$. Figure \ref{Boundaries2}(a)-(b) plots $\sigma_H$ for these two systems as a function of the disorder patch size $L$.  Regions of non-trivial insulating phases are marked in colors, and show quantized integer valued conductivity in units of $e^2/h$.  In between topological phase transitions the system becomes qualitatively gapless and $\sigma_H$ deviates from its quantized values.  Figure \ref{Boundaries2}(c)-(d) plot the local Hall conductivity for various $L$, labeled in 3(a) and 3(b) by roman numerals and gray white circles.  For each choice of $L$, the black dashed lines in the local Hall conductivity plots trace the $g(\bm{r}) = 0.5$ contour. Local fluctuations about the mean conductivity occur around the disorder patch.  In insulating systems these fluctuations must integrate to zero causing regions inside $g(\bm{r})\approx 0.5$ to have a certain signed magnitude, while regions $g(\bm{r})	\lesssim 0.5$ to take on the opposite sign (see Fig. \ref{LineCut}).  Near a phase transition these local fluctuations traverse the system size.  They fold onto themselves forcing the system to change its mean value inducing the topological phase transition.

In the presence of multi-patch disorder configurations the phase boundaries between distinct topological phases show similar behavior to Fig \ref{Boundaries1}(b).  We take a multi-patch disorder configuration of the form $\tau^{\sigma\sigma'}(\bm{r}_i)=-\bm{M}\cdot\bm{\sigma}_{\sigma\sigma'}g(\bm{r}_i)$ with $g(\bm{r}_i)=\sum_{j=1}^{N_p} e^{-|\bm{r}_i-\bm{r}_j|^2/2L^2}$. Here $N_p=N_sp$ with $N_s$ the number of sites in the system and $p$ defined as the percent of disorder.  Similar to how the phase boundaries changed with increasing $L$ for systems with a single patch of disorder, the phase boundaries between distinct topological phases of systems with multi-patch disorder configurations change in a similar way as the percent of disorder is increased.

\begin{figure}
 \includegraphics[width=.45\textwidth]{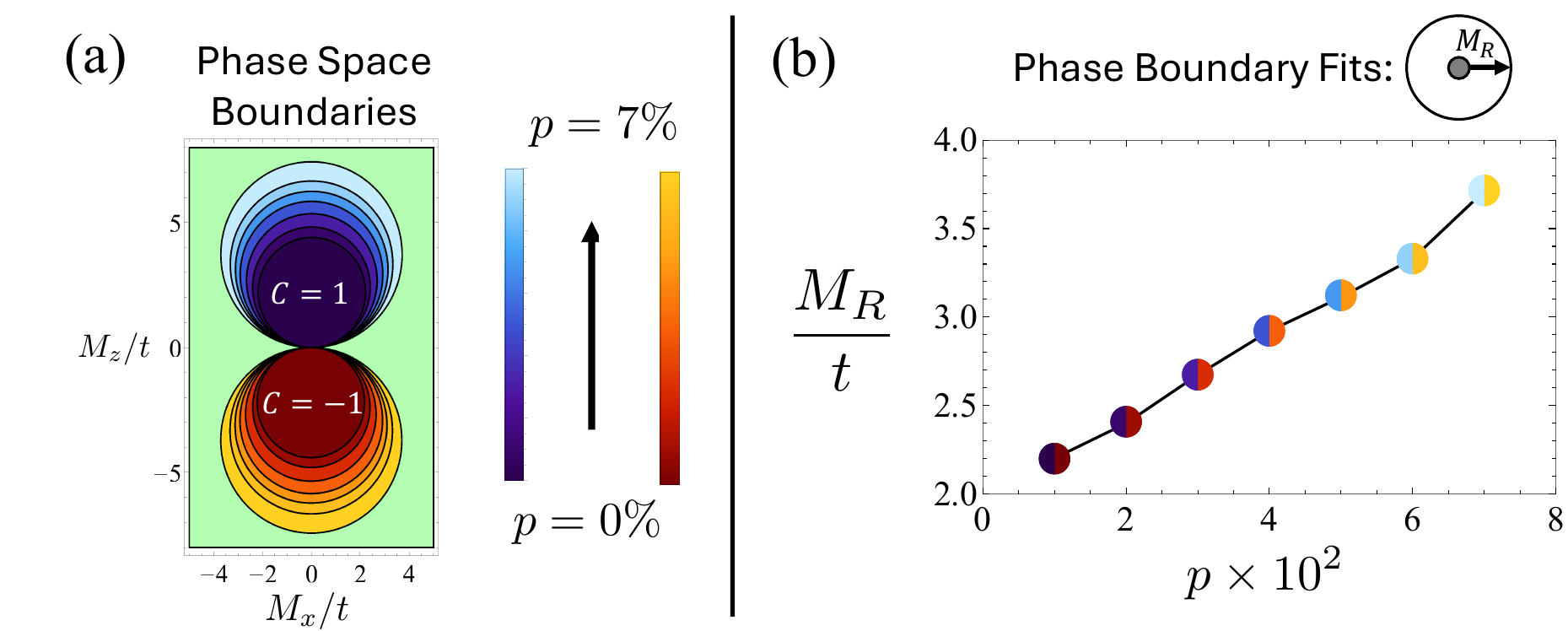}
\caption{{\bf Phase Diagram of Chern Insulator Model for Different Percent of Trivial Potential Disorder.} (a) Phase boundaries in $M_z/M_x$-plane for various amounts of disorder. Shown are the best fits averaged over 25 disorder configurations. Here $N_s=100$ and the defect size is fixed at $L=a$.  (b) We find that the phase space over which the topological insulator exists increases with the percent of disorder; analogous to the behavior for a single patch of disorder shown in Fig.\ref{Boundaries1}.
}
 \label{multiPB}
\end{figure}

Figure \ref{multiPB}(a) shows the disorder profile and corresponding phase boundaries for multi-patch disorder configurations as the percent of disorder increases.  Phase boundaries are circles whose edges are pinned to $\bm{M}=\bm{0}$.  Figure \ref{multiPB}(b) show an almost linear dependence of the radii $M_R$ of different phase boundaries on the percent disorder.

\begin{figure}
 \begin{centering}
 \includegraphics[width=.45\textwidth]{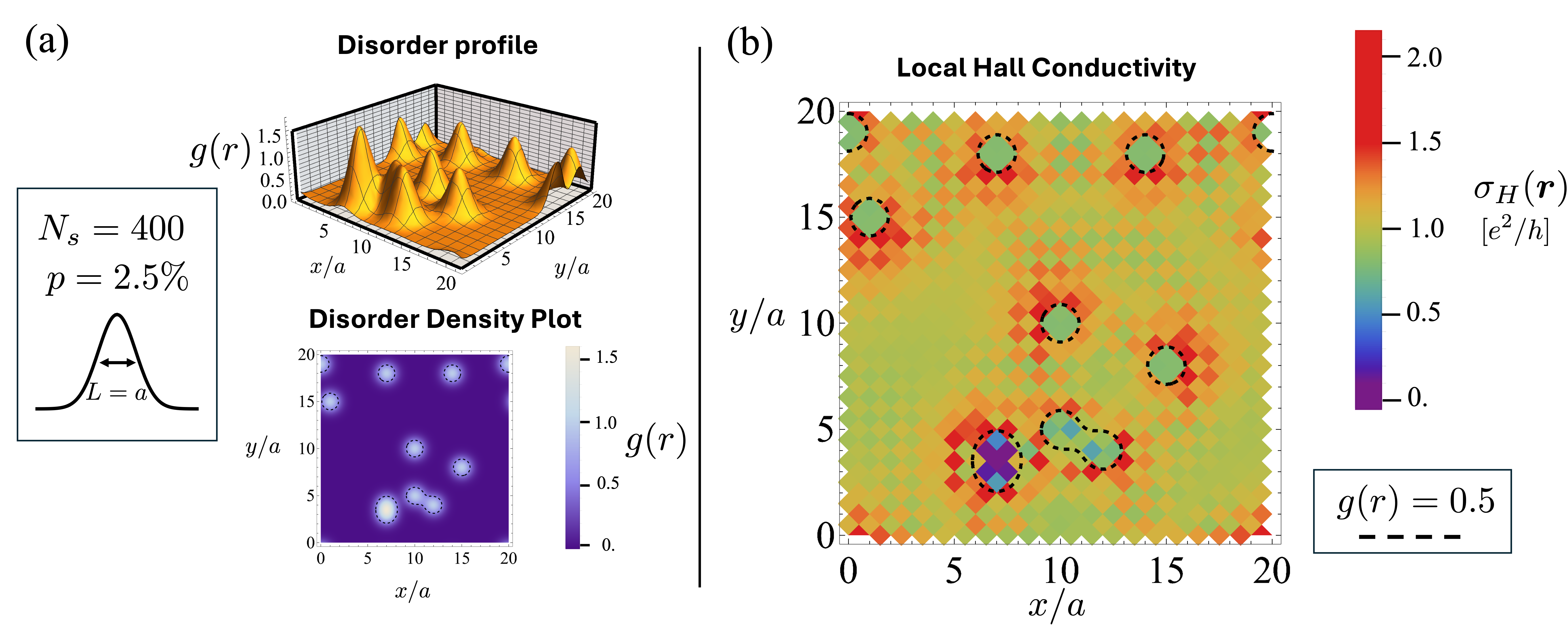}
\caption{{\bf Correlating Disorder Profile and Local Hall conductivity.} (a) Disorder potential profile for $p=2.5\%$, $N_s=400$, and potential disorder with characteristic patch size of $L=a$.  (b) Local Hall conductivity, $\sigma_H(\bm{r})$ for system $\bm{M}=4.15t\bm{\hat{z}}$ with $\sigma_H=e^2/h$ and specific disorder configuration show in (a).  Dashed lines indicate contours where $g(\bm{r})=0.5$ which is when the mass locally is half its unperturbed strength. 
}
 \label{locdis}
 \end{centering}
\end{figure}

An example disorder profile is shown in figure \ref{locdis}(a) for $N_s=400$ and $p=2.5\%$ with patches of size $L=a$.  Figure \ref{locdis}(b) shows the corresponding local conductivity for a system with $\bm{M}=4.15\widehat{\bm{z}}$.  Local Hall conductivity is centered around the global Hall conductivity of $e^2/h$, with large fluctuation shown around the individual disorder patches.  Similar to the local Hall conductivity for a single disorder patch, fluctuations of the Hall conductivity are localized around the center of the disorder patches with the sign of the fluctuations determined approximately by whether the local point of interest sits within or outside the contour $g(\bm{r})=0.5$.

\begin{figure}
 \includegraphics[width=.45\textwidth]{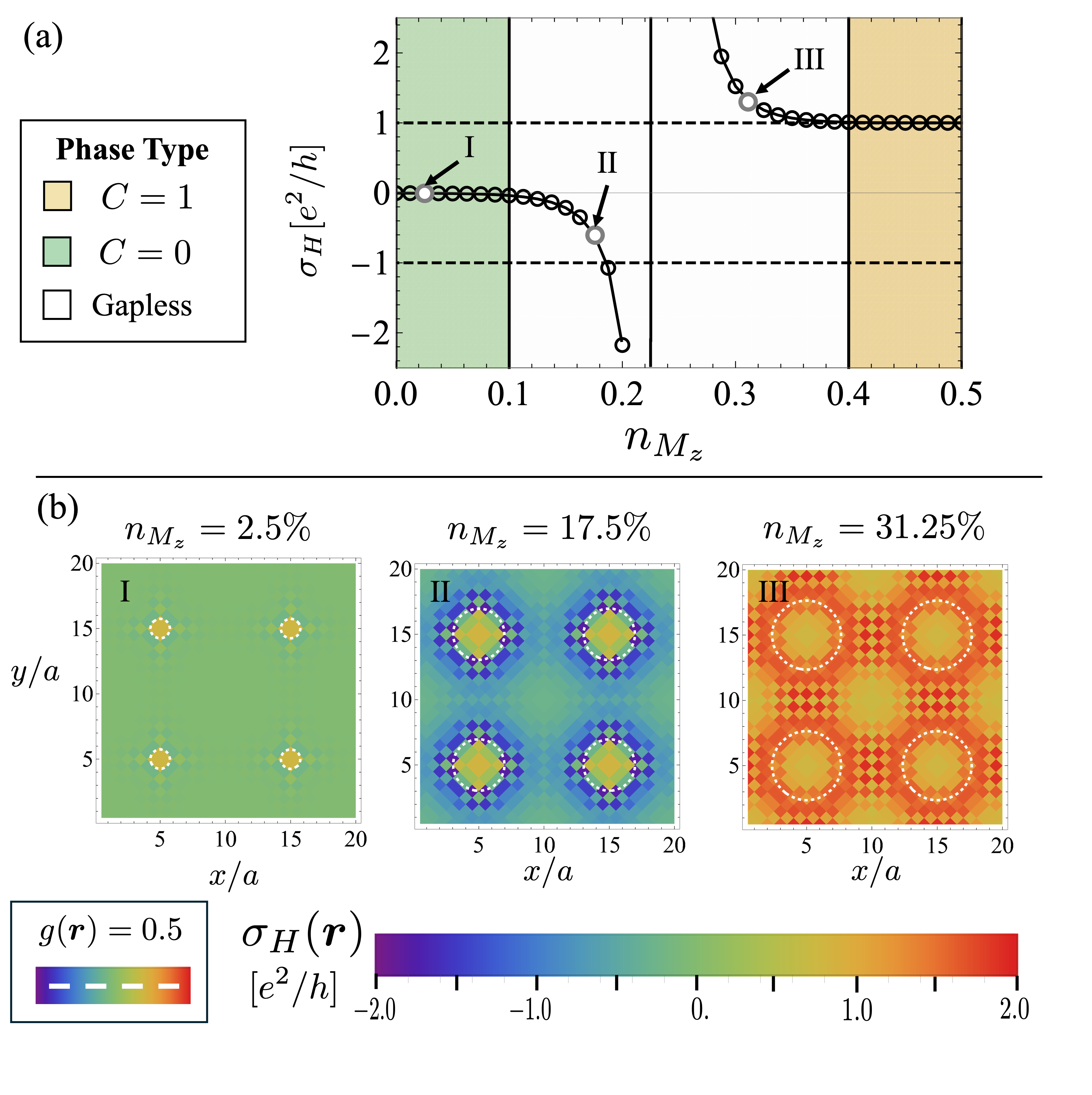}
\caption{{\bf Disorder Patch Correlations.}  (a) Topological phase transition as a function of increasing disorder patch size for a system with $\bm{M}=5.1t\bm{\widehat{z}}$.  (b) Local Hall conductivity around topological phase transition showing fluctuations along paths connecting disorder patches.}
 \label{N4Sweep}
\end{figure}

To investigate the disorder induced topological phase transition in systems with multi-patch disorder configurations we calculate the global and local Hall conductivity for a system with 4 equally spaced patches of disorder.
Figure \ref{N4Sweep} shows the topological phase transition induced in a system with $\bm{M}=5.1t\widehat{\bm{z}}$ when the sizes of the disorder patches are increased.  In figure \ref{N4Sweep}(a) the global Hall conductivity is plotted as a function of the disorder density

\begin{equation}
n_{M_z}=\dfrac{1}{a^2N_s}\int d^2r \, g(\bm{r})    
\end{equation}

\noindent
As the volume of the disorder is increased the system transitions from a trivial $C=0$ to non-trivial $C=1$ state. Figure \ref{N4Sweep}(b) shows the local Hall conductivity for three different $n_{M_z}$.  Across the phase transition fluctuation in the Hall conductivity extend the length of the system and are localized along the shortest paths between disorder patches.  One question is whether the introduction of these multi-patch paths ultimately enhances the non-trivial topological phase space.

To understand the correlation effects between different disorder patches we consider four different disorder configurations (see Fig. \ref{VolLaw}(a)).  We consider systems with 1, 2, 4, and 8 equally space disorder patches.  For each system we calculate the critical $M_z$ along the $M_y=M_x=0$ line for which there is a topological transition.  In figure \ref{VolLaw}(b) we plot the critical $M_z$ value as a function of the topological mass density $n_{M_z}$.  For fixed topological mass density we see that the  the critical $M_z$ grows with the number of disorder patches.  Therefore the addition of multi-patch disorder configurations causes correlations between the disorder patches that lead to fluctuation in the local conductivity along shortest paths between patches that proliferate the phase transition.  If these correlations were absent, the transition would only couple to the zeroth moment of the disorder density and the points plotted in figure \label{VolLaw}(b) would collapse onto a single line.

\begin{figure}
 \includegraphics[width=.45\textwidth]{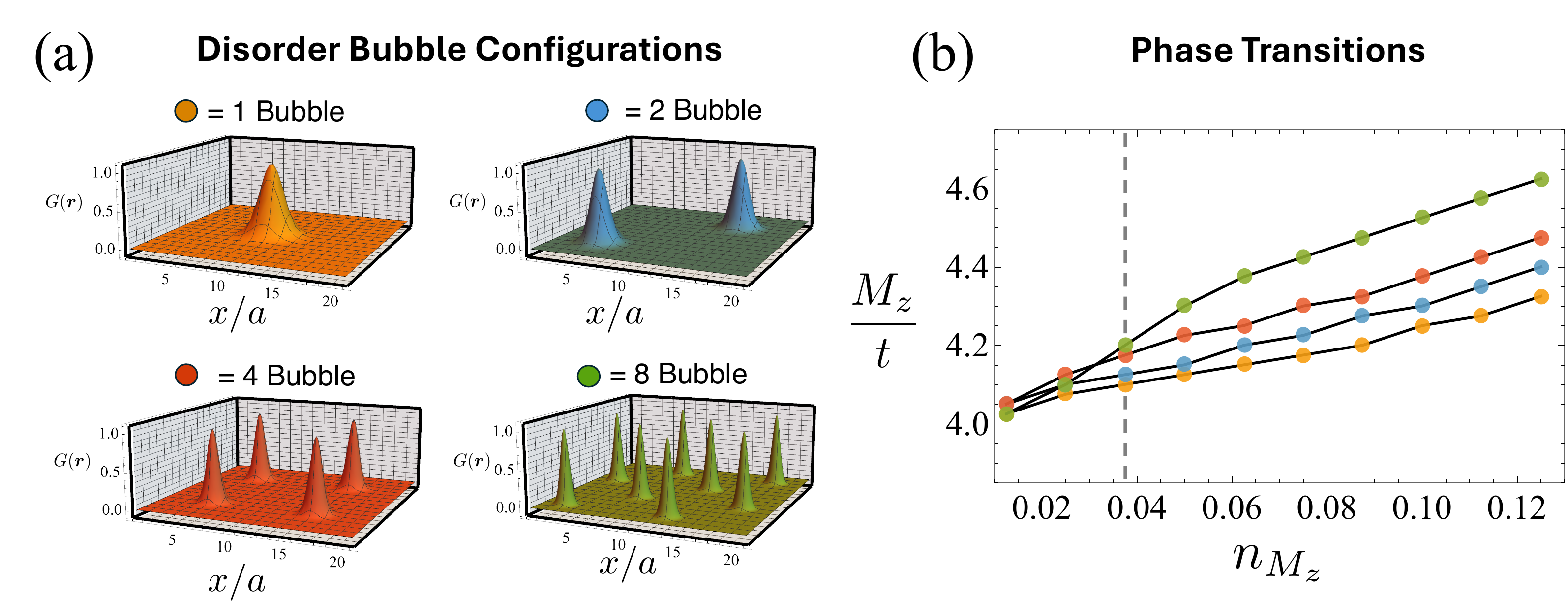}
\caption{{\bf Disorder Correlations.
}  (a) Local distribution of disorder for systems with one, two, four, and eight equally spaced patches.  The global density of each configuration is $n_{M_z}=3.75\%$ (dashed line in (b)). (b) Critical point separating topological and trivial phases along $M_x=M_y=0$ line as a function of $n_{M_z}$. The topological phase exists over a larger phase space whose critical $M_z$ is shifted up to larger values when there are more disorder patches with $n_{M_z}$ held fixed. }
 \label{VolLaw}
\end{figure}

\section{Conclusion}

To understand the distinct topological phase transition induced by our choice of Anderson disorder we can look toward the phase boundaries depicted in Fig \ref{Boundaries1}.  In the absence of disorder a trivial insulator has a large mass $\bm{M}$.  Doping the system with trivial semi-metallic patches drives the system to an effectively smaller mass pushing the system into a Chern insulating phase.  This is made evident in the local picture of transport, in particular, for a system with a single patch of disorder where there is a contour that separates regions of positive and negative local Hall conductivity.  As shown in Fig. \ref{Boundaries2} topological transitions occur when the contour separates regions of almost equal area.  However, this naive picture would suggest that the size of the patches of disorder completely determines the topological transition.  As we have shown in Fig. \ref{VolLaw}, correlations between patches are crucial in predicting where the topological transition occurs.  By looking at different disorder patch configurations we have shown that it is advantageous to breakup a large patch of disorder into small patches to stabilize a topological insulating state.

The local Hall conductivity that we propose here has some advantages over the other local topological markers. (i) It is most closely related to an experimental observable; (ii) it can be calculated in the absence of translation symmetry for finite systems with a boundary or systems with periodic boundary conditions. In both contexts the current operator used in the local Hall conductivity is well defined rather than the use of position operators that are ill-defined for periodic systems; (iii) It allows a simple calculation of the bulk Hall conductivity via integration of the local quantity. 

The local picture of transport currents contains information about the local translation symmetry breaking of a disordered crystal.  Here we have shown how the local Hall conductivity can be used to help determine effective ways to induce topological transitions in trivial magnetic insulators.  The topological Anderson insulator has been experimentally realized in the context of 1D atomic wires \cite{meier2018observation} and photonic coupled waveguides \cite{stutzer2018photonic}.  While scanning superconducting interference devices (SQUID) can measure the local magnetization of a sample with resolution $\sim 0.1\text{--}10\mu m$ \cite{persky2022studying}, here topological information is more readily available in the local Hall conductivity whose relation to the Chern number is more direct.  We propose that the Streda formula, which has been used as a measure of the global Hall conductance, can be applied locally as a scanning probe to extract the local Hall conductance along a cut across the system to determine the topological nature of the state, as depicted in Fig. ~4. 
Likewise scanning microwave impedance measurements (MIM) determine the local longitudinal conductivity with resolution $\gtrsim 1nm$ \cite{tami2025scanning}, whereas here topological information is in the antisymmetric part of the conductivity tensor.  Both techniques also being limited by their spatial to signal resolution contrast.  The best ways to measure directly the local Hall conductivity at length scales relevant to most disorder potentials is still an open question.  

In the context of magnetic metals, rather than insulating systems, the Hall conductivity has an anomalous contribution proportional to the magnetization of the system.  In the presence of translation symmetry breaking these systems in addition to the anomalous Hall effect are seen to have a topological Hall effect proportional to the skyrmion density \cite{nagaosa2013topological,lee2009unusual,neubauer2009topological,kanazawa2011large,li2013robust,gallagher2017robust,ahmed2018chiral,ahmed2019spin,shao2019topological}.  In principle the formalism introduced above could determine the local Hall currents associated with such structured defects.

Our theory is developed in a full real-space picture of transport, while in fully periodic systems the conservation of the crystal momentum makes momentum space a more direct way to understand the Hall conductivity.  For disorder distributions that change smoothly on a scale $\gg a$ a local picture of transport can be determined in the semiclassical framework where momentum and position space variables are treated on equal footing.  This framework has been used to study the Hall currents in metals which can be described by contributions deriving from both real and momentum space Berry curvatures \cite{ye1999berry,bruno2004topological,nagaosa2012gauge,verma2022unified}.  An open question is how these curvatures manifest in the fully quantum framework described here.  We leave these questions for future work.

\medskip
\noindent
{\bf Acknowledgments:} ZA and NT acknowledge the National Science Foundation Division of Mathematical Sciences for the grant award NSF-DMS 2138905.

\bibliography{Bibliography}

\clearpage 

\onecolumngrid

\appendix

\section{Local Conductivity}\label{IdLoc}

Here we are interested in the response of a system to a spatially homogeneous electric field at zero frequency.  As such we may take as our gauge fields $\bm{A}_{ij}(t)=\bm{A}(t)$ independent of the bond index.  To extract the local conductivity we can expand the local current to linear order in the gauge field and Fourier transform in time:

\begin{equation}
\bm{j}^{(1)}_{ij}(\omega)=\sum_{n,m}\rho^{(1)}_{nm}(\omega)\bm{j}_{mn}^{(0)}(\bm{r}_{ij},0)+\rho^{(0)}_{nm}(0)\bm{j}_{mn}^{(1)}(\bm{r}_{ij},\omega)
\label{currloc}
\end{equation}

\noindent
Note that we are looking at the local current in response to a homogeneous field such that the matrix elements of the density matrix remain spatially homogeneous, whereas the local current operator depends on position through its bond indices.  From $\widehat{\bm{j}}(\bm{r}_{pq},t)= -\bm{\nabla}_{\bm{A}(\bm{r}_{pq},t)} \widehat{H}(t)/V$ we find

\begin{align}
\bm{j}_{nm}^{(0)}(\bm{r}_{ij},\omega)&=- \delta_{\omega,0}\dfrac{1}{V}\Psi^*_n(\bm{r}_i)\bra{\bm{r}_{i}}  \bm{\nabla}_{\bm{A}(t)}\widehat{H}(t)\ket{\bm{r}_j}\Psi_m(\bm{r}_j) \nonumber \\
    \bm{j}_{nm}^{(1)}(\bm{r}_{ij},\omega)&=-\dfrac{1}{V}\sum_q\bigg(\Psi^*_n(\bm{r}_i)\bra{\bm{r}_{i}} \bm{\nabla}_{\bm{A}(t)}\partial_{A_q(t)}\widehat{H}(t)\ket{\bm{r}_j}\Psi_m(\bm{r}_j)\bigg)A_q(\omega)
\label{loccurrop}
\end{align}

\noindent
where $\ket{\bm{r}_i}$ is the local basis state $\ket{\bm{r}_i}=\widehat{c}^\dagger_{\bm{r}_i}\ket{0}$. For simplicity in what follows we define a spatially resolved inverse mass tensor

\begin{equation}
(M^{-1})_{ij}^{pq}=\bra{\bm{r}_{i}} \partial_{A_p(t)}\partial_{A_q(t)}\widehat{H}(t)\ket{\bm{r}_j}/e^2
\end{equation}

\noindent
and local velocity operator

\begin{equation}
    \tilde{\bm{v}}_{ij}=\dfrac{1}{e}\bra{\bm{r}_{i}}  \bm{\nabla}_{\bm{A}(t)}\widehat{H}(t)\ket{\bm{r}_j}
\end{equation}

To calculate the local Hall conductivity of a system it is advantageous to separate the zero frequency pole in the conductivity from its other contributions.  To accomplish this task we note that the analogous express to equation \ref{CId} in the main text for the local current operators may be expressed as

\begin{align}
   & \partial_{A_q(t)}\bigg(\braket{n_{\bm{A}(t)}|\bm{r}_i}\bra{\bm{r}_{i}}  \partial_{A_p(t)}\widehat{H}(t)\ket{\bm{r}_j}\braket{\bm{r}_j|n_{\bm{A}(t)}}\bigg)/e^2\nonumber \\
   & =\dfrac{1}{e}\bigg(\partial_{A_q(t)}\braket{n_{\bm{A}(t)}|\bm{r}_i}\bigg)\tilde{v}^p_{ij}\Psi_n(\bm{r}_j) 
    +\dfrac{1}{e}\Psi_n^*(\bm{r}_i)\tilde{v}^p_{ij}\bigg(\partial_{A_q(t)}\braket{\bm{r}_j|n_{\bm{A}(t)}}\bigg)+\Psi_n^*(\bm{r}_i)M_{ij}^{pq}\Psi_n(\bm{r}_j) \nonumber \\
   & =\dfrac{1}{e}\sum_m\bigg[\bigg(\partial_{A_q(t)}\braket{n_{\bm{A}(t)}|m}\bigg)\Psi^*_m(\bm{r}_i)\tilde{v}^p_{ij}\Psi_n(\bm{r}_j)
    +\Psi_n^*(\bm{r}_i)\tilde{v}^p_{ij}\Psi_m(\bm{r}_j)\bigg(\partial_{A_q(t)}\braket{m|n_{\bm{A}(t)}}\bigg)\bigg]
    +\Psi_n^*(\bm{r}_i)M_{ij}^{pq}\Psi_n(\bm{r}_j)
    \label{locid}
\end{align}

The contributions to the current derive from the two terms in equation \eqref{currloc}.  The equations for $\rho_{nm}^{(1)}(\omega)$ and $\rho_{nm}^{(0)}(0)$ remain unchanged as the type of perturbing electric field remains spatially homogenous.  Using equation \eqref{denmat} in the main text and equations \eqref{loccurrop} and \eqref{locid} we may write the diamagnetic contribution to the local current as

\begin{align}
    \bm{j}^D_{ij}(\omega)\cdot\widehat{\bm{r}}_p&=\sum_{n,m}\rho^{(0)}_{nm}(0)\bm{j}_{mn}^{(1)}(\bm{r}_{ij},\omega)\cdot\widehat{\bm{r}}_p \nonumber \\
    &=-\dfrac{e^2}{V}\sum_{n,m,q}\bigg[\delta_{nm}P_n\partial_{A_q(t)}\bigg(\braket{n_{\bm{A}(t)}|\bm{r}_i}\tilde{v}_{ij}^p\braket{\bm{r}_j|n_{\bm{A}(t)}}\bigg)/e-(P_n-P_m)\dfrac{v^q_{nm}\Psi_m^*(\bm{r}_i)\tilde{v}_{ij}^p\Psi_n(\bm{r}_j)}{E_n-E_m}\bigg]A_q(\omega)
\end{align}

\noindent
and the paramagnetic contribution to the local current as 

\begin{equation}
    \bm{j}^P_{ij}(\omega)\cdot\widehat{\bm{r}}_p=\dfrac{1}{V}\sum_{n,m}\rho^{(1)}_{nm}(0)\bm{j}_{mn}^{(0)}(\bm{r}_{ij},\omega)\cdot\widehat{\bm{r}}_p=-\dfrac{e^2}{V}\sum_{n,m,q}\dfrac{(P_n-P_m)v^q_{nm}\Psi_m^*(\bm{r}_i)\tilde{v}_{ij}^p\Psi_n(\bm{r}_j)}{E_n-E_m+\hbar\omega} A_q(\omega) 
\end{equation}

\noindent
This allows us to write the total local current as

\begin{align}
    \bm{j}_{ij}^{(1)}(\omega)\cdot\widehat{\bm{r}}_p=&-\dfrac{e^2}{V}\sum_{n,m,q}\bigg[\delta_{nm}P_n\partial_{A_q(t)}\bigg(\braket{n_{\bm{A}(t)}|\bm{r}_i}\tilde{v}_{ij}^p\braket{\bm{r}_j|n_{\bm{A}(t)}}\bigg)/e-(P_n-P_m)\dfrac{v^q_{nm}\Psi_m^*(\bm{r}_i)\tilde{v}_{ij}^p\Psi_n(\bm{r}_j)}{E_n-E_m}+\nonumber \\
    &+\dfrac{P_n-P_m}{E_n-E_m+\hbar\omega}v^q_{nm}\Psi_m^*(\bm{r}_i)\tilde{v}_{ij}^p\Psi_n(\bm{r}_j) \bigg] A_q(\omega) \nonumber \\
    =&-\dfrac{e^2}{V}\sum_{n,m,q}\bigg[\delta_{nm}P_n\partial_{A_q(t)}\bigg(\braket{n_{\bm{A}(t)}|\bm{r}_i}\tilde{v}_{ij}^p\braket{\bm{r}_j|n_{\bm{A}(t)}}\bigg)/e-\dfrac{ (P_n-P_m)v^q_{nm}\Psi_m^*(\bm{r}_i)\tilde{v}_{ij}^p\Psi_n(\bm{r}_j)}{(E_n-E_m)(E_n-E_m+\hbar\omega)}\,\hbar\omega \bigg] A_q(\omega) \nonumber \\
    =&-\dfrac{e^2}{V}\sum_{n,m,q}\bigg[\dfrac{\delta_{nm}P_n}{i\omega}\partial_{A_q(t)}\bigg(\braket{n_{\bm{A}(t)}|\bm{r}_i}\tilde{v}_{ij}^p\braket{\bm{r}_j|n_{\bm{A}(t)}}\bigg)/e+i\hbar\dfrac{(P_n-P_m) v^q_{nm}\Psi_m^*(\bm{r}_i)\tilde{v}_{ij}^p\Psi_n(\bm{r}_j)}{(E_n-E_m)(E_n-E_m+\hbar\omega)} \bigg] E_q(\omega)    
\end{align}

\noindent
Using this expression the local bond conductivity is given by

\begin{equation}
\sigma_{pq}(\bm{r}_{ij},\omega)=-\dfrac{e^2}{V}\sum_{n,m,q}\bigg[\dfrac{\delta_{nm}P_n}{i\omega}\partial_{A_q(t)}\bigg(\braket{n_{\bm{A}(t)}|\bm{r}_i}\tilde{v}_{ij}^p\braket{\bm{r}_j|n_{\bm{A}(t)}}\bigg)/e+i\hbar\dfrac{(P_n-P_m) v^q_{nm}\Psi_m^*(\bm{r}_i)\tilde{v}_{ij}^p\Psi_n(\bm{r}_j)}{(E_n-E_m)(E_n-E_m+\hbar\omega)} \bigg]
\label{bondcur}
\end{equation}

\noindent
Note that $\sum_{ij}\sigma_{pq}(\bm{r}_{ij},\omega)=\sigma_{pq}(\omega)$.  We thus define the local conductivity as

\begin{equation}
\sigma^{\text{loc}}_{pq}(\bm{r}_{ij},\omega)\equiv\sigma_{pq}(\bm{r}_{ij},\omega)+\sigma_{pq}(\bm{r}_{ji},\omega)
\end{equation}

\noindent
Using the identities above we may rewrite the first term in equation \eqref{bondcur} as

\begin{equation}
\sigma_{pq}^D(\bm{r}_{ij},\omega)=-\dfrac{1}{i\omega}\dfrac{e^2}{V}\sum_{n,m,q}\bigg[i(P_n-P_m)\dfrac{v^q_{nm}\Psi_m^*(\bm{r}_i)\tilde{v}_{ij}^p\Psi_n(\bm{r}_j)}{E_n-E_m}+
    \delta_{nm} P_n\Psi_n^*(\bm{r}_i)(M^{-1})_{ij}^{pq}\Psi_n(\bm{r}_j)\bigg]
\end{equation}

\noindent
defining $\sigma_{pq}^D(\bm{r}_{ij},\omega)$ as the local diamagnetic contribution to the current.

\end{document}